# Dislocation interaction with C in α-Fe: a comparison between atomic simulations and elasticity theory.

Emmanuel Clouet[1,2], Sébastien Garruchet[3], Hoang Nguyen[2], Michel Perez[3], Charlotte S. Becquart[2]

[1]Service de Recherches de Métallurgie Physique, CEA/Saclay, 91191 Gif-sur-Yvette, France

[2]Laboratoire de Métallurgie Physique et Génie des Matériaux, UMR CNRS 8517, École Nationale Supérieure de Chimie de Lille, Bât. C6, 59655 Villeneuve d'Ascq cedex, France

[3]MATEIS, UMR CNRS 5510, Université de Lyon, INSA-Lyon, 69621 Villeurbanne, France

**Abstract**

The interaction of C atoms with a screw and an edge dislocation is modelled at an atomic scale using an empirical Fe-C interatomic potential based on the Embedded Atom Method (EAM) and molecular statics simulations. Results of atomic simulations are compared with predictions of elasticity theory. It is shown that a quantitative agreement can be obtained between both modelling techniques as long as anisotropic elastic calculations are performed and both the dilatation and the tetragonal distortion induced by the C interstitial are considered. Using isotropic elasticity allows to predict the main trends of the interaction and considering only the interstitial dilatation will lead to a wrong interaction.

**Keywords**: Fe-C alloys, screw dislocation, edge dislocation, anisotropic elasticity, Cottrell atmospheres, binding energy.

**Corresponding author:** Charlotte Becquart.
Tel: (33) 3 20 43 49 44
Email : charlotte.becquart@univ-lille1.fr

# Introduction

Interactions between interstitial solute atoms and dislocations drive many mechanical properties of steels. Indeed, carbon or nitrogen atoms in body centred cubic (bcc) iron tend to segregate on dislocations and to form Cottrell atmospheres [1]. Once these atmospheres appeared, an extra force is needed to unpin the dislocations. These pinning/unpinning processes lead to a subsequent higher yield stress and to mechanical instabilities (Lüders' bands) [2] that are a serious hindrance to manufacture. Moreover, interaction between interstitial atoms and dislocations also limits the life span of many metallic components (strain ageing effects). Static strain ageing implies the formation of Cottrell atmospheres whereas dynamic strain ageing occurs at high temperature and involves a competitive motion of dislocations and interstitial atoms [1]. To prevent any pinning of dislocations by interstitial atoms ultra-low carbon or interstitial free steels have been designed by removing interstitial atoms out of the solid solution: either by involving them within precipitates (TiN, TiC,…) or by making them interact with other substitutional atoms (Mn, Cr). In some cases, interaction between dislocations and solute atoms is also responsible of heterogeneous precipitation (*e.g.* NbC in iron [3, 4]). This interaction may cause a segregation of solute atoms on the dislocation and, locally, the supersaturation could become high enough for precipitates to nucleate. The estimation of the dislocations – interstitial atoms interaction is thus important to understand and model both the flow behaviour of steels and the first stages of heterogeneous precipitation.

Various experimental techniques give insight on this interaction. Using thermoelectric power, resistivity measurements or mechanical spectroscopy [5-7], it is possible to obtain a value for the segregation energy. One can also use three-dimensional atom probe tomography to directly image the Cottrell atmospheres decorating the dislocations [8-10].

Atomic simulations are also a good tool to study this interaction. For instance, using different interatomic potentials, several authors [11-15] obtained a segregation energy for C atoms on dislocations in reasonable agreement with the ones deduced from experiments. It is also possible to directly simulate the effect of interstitial impurity on the dislocation glide properties via molecular dynamics [14]. But, despite their capacity, atomic simulations have severe drawbacks as they do not allow to model more than a few dislocations (usually one or two), whereas the flow behaviour of a metal is mainly controlled by the collective evolution of the whole dislocation population. Moreover, the timescale that can be simulated with molecular dynamics is not compatible with the one corresponding to solute diffusion, preventing any direct simulation at the atomic scale of the Cottrell atmosphere formation. Therefore whether one is interested in the increase of the yield stress associated with the pinning / unpinning process of the dislocations from their Cottrell atmospheres, or in the segregation kinetics of solute interstitials on dislocations, atomic simulations are not sufficient and a modelling at a higher scale is also needed. It is thus necessary to describe the interaction between interstitials and dislocations not only at the atomic scale as can be done with ab-initio or empirical potentials, but also at a mesoscopic scale compatible with other simulation and modelling tools.

For that purpose, elasticity theory has been widely used: from the calculation of the elastic field around a dislocation [16] and the elastic distortion of the host lattice due to an interstitial atom [17], this theory can estimate the interaction energy between both defects. Cottrell and Bilby [1] first estimated this energy considering the elastic interaction between the pressure created by a dislocation and the relaxation volume of C interstitials in iron (the size interaction). This calculation was then improved by Cochardt *et al.* [18] so as to consider not only the dilatation but also the shear strain associated with interstitials (the shape interaction). All these calculations were performed using isotropic elasticity to get the dislocation stress field, although iron is anisotropic due to the cubic symmetry of its lattice. Douthwaite and Evans [19] thus repeated the calculations of Cochardt *et al.* using anisotropic elasticity to obtain the dislocation elastic field. With their work, the modelling within linear elasticity theory of the interaction between dislocations and C interstitials in iron was then complete.

The simplicity and versatility of elasticity theory is at the origin of its wide success. However, this theory cannot quantify the interaction when the interstitial atoms lie in the dislocation core. The question of the validity of elasticity theory remains open: what is the minimal separation distance between both defects so as to trust linear elasticity theory? Moreover, most of elastic calculations assume isotropy and only consider the interstitial dilatation like in the original work of Cottrell and Bilby [1]. These simplifying assumptions need to be checked as they may be wrong. In particular, the work of Cochardt *et al.* [18] showed that the shape interaction has to be considered to account for the existence of Cottrell atmospheres around screw dislocations.

One way to assess the validity of elasticity theory is by means of atomic simulations so as to compare the interaction energies between interstitial atoms and dislocations given by both methods. Such an approach has already been used to study the interaction between vacancies and dislocations in face centred cubic (fcc) metals [20]. In that case, isotropic linear elasticity led to quantitative predictions of the interaction energy as soon as the vacancy was further than a few atomic distances (~10 Å) from the dislocations. Studying now the dislocation interaction with an interstitial atom will extend this conclusion to the case where the point defect not only acts as a dilatation centre but also causes a shear. Moreover, choosing a bcc host lattice allows to avoid the complexity of dislocations in fcc lattice related to the stacking fault which cannot be easily modelled within elasticity theory.

In the work described in the present article, we use molecular statics (MS) to investigate the interaction between a carbon atom and screw or edge dislocations in bcc iron at an atomic scale. These simulations rely on a recently developed empirical interatomic potential for Fe-C [13, 21]. Interaction energies computed with MS are then compared with elasticity theory for a wide number of C atom positions around the dislocation. This allows us to examine the validity of the different approximations that can be made in the elastic calculation.

# I Atomic simulations

## I.1 Interatomic potential

The FeC interatomic potential used in this work was built according to the Embedded-Atom Method [22,23]. The pure Fe part was developed by Mendelev and co-workers [24,25] and the Fe-C interaction by J.M. Raulot [13]. This potential was fitted on experimental and ab initio data. It should be stressed that care was taken in the fitting procedure of the Fe part so as to ensure a reasonable description of atomic interactions at small separation distances (see Ref. 24 for a detailed description). As a consequence, it leads to a core structure for the screw dislocation in pure α-Fe in agreement with ab initio calculations [26-28]: the core is compact rather than exhibiting a degenerated core, *i.e.* asymmetrically spread in the three {1 1 0} of the [1 1 1] zone as it is very often predicted by empirical potentials for bcc Fe. With this potential, the most stable configuration for C in interstitial configuration is the octahedral site, in agreement with experimental observations [29]. C diffusion in α-Fe and the evolution of the lattice parameter versus C content are also in good agreement with the experimental data [13].

## I.2 Introduction of the dislocation in the simulation cell

The Burgers vector of both the edge and screw dislocations considered is $b = a_0/2$ [111], where $a_0 = 2.8553$ Å is Fe lattice parameter as given by the atomic potential, and their glide plane is a {110} plane. These dislocations are the most commonly observed in iron. For the edge dislocation, the dislocation line is in the $[1\bar{2}1]$ direction, whereas for the screw dislocation, the dislocation line is in the [111] direction.

The dislocations were created at the centre of the simulation box by displacing atoms according to anisotropic elasticity theory of straight line defects [30-32], *i.e.* by applying to each atom the displacement corresponding to the dislocation Volterra elastic field. In the case of the screw dislocation, some analytical expressions of this elastic field have been developed [33,34] and can be used instead of the general sextic formalism [30-32]. All the atoms of the lattice except for those situated in an outside layer 8 Å thick (twice the potential cut-off) were then allowed to relax so as to minimize the simulation box energy as given by the atomic potential. The atoms in the outside layer were fixed in the position given by anisotropic elasticity.

Depending on the sign of the Burgers vector, two possible configurations of the screw dislocation can be obtained most of the time referred to as "hard" and "soft" [35]. In this work, the most stable configuration, (*i.e.* "soft"), has been investigated.

**I.3 Detection of the dislocation location**

Due to its high Peierls stress, the screw dislocation stays in its initial location. However this is not the case for the edge dislocation. When a carbon atom lies close to the edge dislocation core, the interaction between the two defects becomes so strong that the dislocation moves along its glide plane. In this case, it is important to locate precisely the final position of the dislocation core. A convenient way to do so is to compare the atomic simulations with the predictions of a Peirls-Nabarro model [36]. The dislocation line is detected by calculating the disregistry function of the atoms, that is the displacement difference $D(x) = u_{above}(x) - u_{below}(x)$ between the atoms in the two adjacent planes [111] above and below the slip plane. The derivative of this function is then computed giving thus the Burgers vector density distribution:

$$\rho(x) = \frac{dD(x)}{dx},$$

where $x$ is the coordinate along the gliding direction.

Using the Peierls-Nabarro model, these two functions get the following simple expressions [16]:

$$D(x) = \frac{b}{\pi} \tan^{-1}\left(\frac{x}{\xi}\right) + \frac{b}{2}$$

and

$$\rho(x) = \frac{dD(x)}{dx} = \frac{b\xi}{\pi(\xi^2 + x^2)},$$

where $\xi = d/[2(1-\nu)]$ can be considered as the dislocation half width, $d = \sqrt{2}a_0$ is the interplanar spacing in the direction perpendicular to the $(\bar{1}01)$ glide plane and $\nu$ is Poisson ratio.

Figures 1.a and 1.b represent the disregistry function and the Burgers vector density distribution for the edge dislocation in α-Fe observed in our atomic simulations after the dislocation has been created using anisotropic elasticity and the atomic positions have been relaxed according to the interatomic potential. The disregistry function and the corresponding distribution predicted by the Peierls-Nabarro model are also presented for comparison. Despite some discrepancies, both density distributions are maximum for the same position. This position corresponds to the dislocation core in the Peierls-Nabarro model. Therefore the location of the edge dislocation will be associated with the maximum of this density distribution in our simulations.

**I.4 C interstitials in interaction with the dislocation**

In α-Fe, interstitial solute carbon atoms are found in octahedral sites. This kind of site has two first neighbours oriented along a ⟨100⟩ type direction and four second nearest neighbour lying along ⟨110⟩ directions in the plane perpendicular to the direction containing the two first neighbours. The site has full tetragonal symmetry [(4/m)(2/m)(2/m)]. The two first neighbours ⟨100⟩ direction corresponding to the tetragonal axis can be referred to as "the site orientation". There are thus three variants of these octahedral sites: the [100], [010] and [001] sites which are energetically equivalent in a stress free state.

The binding energy between the C atom and the dislocation has been determined employing molecular statics. The following definition has been used:

$$E^{bind}(disloc, octa) = [E(disloc) + E(octa)] - [E(disloc + octa) + E(ref)] \quad \text{(eq. 1)},$$

where $E(disloc)$ is the total energy of the system containing one dislocation, $E(octa)$ the total energy of the system containing one carbon atom in an octahedral site, $E(ref)$ the total energy of the perfect lattice, and $E(disloc + octa)$ the total energy of the system containing both one dislocation and one carbon atom. With such a definition, a positive binding energy indicates attraction between the interstitial and the dislocation.

The C atom has been introduced in the neighbourhood of the dislocation line, in octahedral sites, after an initial relaxation of the dislocation. Then the simulation box containing the carbon atom and the dislocation has been relaxed again.

**I.5 Results of the atomic simulations**

The results for the screw dislocation are presented in figure 2 on which the C binding energy has been represented for one variant. Due to the 3-fold symmetry obeyed both by the bcc lattice (Fig. 2a) and the dislocation, the binding energy for other variants of C atoms can be simply obtained by a ± 2π/3 rotation of figure 2b. We checked in our atomic simulations that this 3-fold symmetry holds for the C – screw dislocation binding energy. The most stable configuration for C atom does not correspond to the closest one to the dislocation core, as this position is highly unstable, but is situated on the next neighbour shell.

For the edge dislocation, as expected, the binding energy is largely positive on the tension side near the core which corresponds to a strong attraction, and slightly negative on the compression side which corresponds to a slight repulsion. The maximum carbon-edge dislocation interaction energy arises for carbon atoms located in octahedral sites with tetragonal distortion axis in the direction [100] or [001], in the glide plane, very close to the dislocation core (±0.3 Å according to the site type). This value, $E_b$ = 0.66 eV, is very close to that of 0.68 eV found recently by Tapasa et al. [14] or the value of 0.7 eV obtained by de Hosson [12]. The second highest energy value, $E_b$ = 0.47 eV, is again for octahedral sites with tetragonal distortion axis in the direction [100] or [001] but is very close to the maximum value for the [010] site, $E_b$ = 0.42 eV, for a carbon in the site just below the glide plane.

Table I summarizes the maximum binding values obtained in this work and compares them with experimental data as well as with the results of other atomic simulations [11,12,14,15]. An interesting point is that the maximum binding energy for the edge dislocation (0.66 eV) is approximately 60% higher than the one for the screw dislocation (0.41 eV). It is worth noting that the different atomic simulations (cf. Tab. 1) agree on this stronger binding. Therefore, one can expect that C will pin edge dislocations more efficiently than screw dislocations. Nevertheless, the pinning of screw dislocations by C atoms remains effective. The values obtained in the present work are compatible with the ones deduced from experiments. One could not go further in the comparison as experiments do not allow differentiating between edge and screw dislocations. Moreover the different experimental data are quite scattered even when they are obtained with the same techniques. The values obtained from atomic simulations by different authors show some variations too. It illustrates the sensitivity of the C-dislocation binding energy to the empirical potential that was used. Even the most recent studies [14,15], including this work, lead to different values although the potentials used give the same elastic constants for Fe. This clearly shows that the maximal carbon – dislocation binding energy cannot be directly deduced from elasticity. For the sake of completeness, we nevertheless include in Table 1 binding energies given by elastic calculations [18,19]. As we shall see in the next section, the cut-off distance used in these calculations may be too small for elasticity to apply. But, it is clear that the values predicted by elasticity are surprisingly reasonable compared to the experimental ones and to the ones deduced from our atomic simulations.

## II Elasticity theory
### II.1 Point defect description

Elasticity theory assumes a continuum description, within which a point defect can be modelled by a tensor $P_{ij}$ corresponding to the moments of an equilibrated point force distribution [37]. The tensor $P_{ij}$, usually called "elastic dipole", can be directly deduced from simple atomic simulations. To do so, one can consider a periodic simulation box having a volume $V$ and containing only one point defect. Such a simulation box can minimize its energy by taking a homogeneous strain $\varepsilon_{ij}$. The contribution $E_\varepsilon$ of such a strain to the elastic energy will be composed of its self energy and its interaction with the force moment tensors [37],

$$E_\varepsilon = \frac{1}{2} V C_{ijkl} \varepsilon_{ij} \varepsilon_{kl} - P_{ij} \varepsilon_{ij}, \qquad \text{(Eq. 2)}$$

where $C_{ijkl}$ are the elastic constants of the host crystal. If the system is free to relax, it will adopt the strain that minimizes its energy. The elastic energy given by Eq. 2 is minimal when the following relation between the homogeneous strain and the elastic dipoles is verified,

$$P_{ij} = V C_{ijkl} \varepsilon_{kl}. \qquad \text{(Eq. 3)}$$

Doing atomic simulations where the system is allowed to relax the atomic coordinates as well as the coordinates of the periodicity vectors, i.e. relaxation under no external stress, one gets direct access to the homogeneous strain $\varepsilon_{ij}$ induced by the point defect on the simulation box and therefore on the tensor $P_{ij}$ used to model this point defect.

Usually, it is more convenient to perform atomic simulations where only the atomic coordinates are relaxed whereas the periodicity vectors are kept fixed. A simulation box containing a point defect will therefore develop a stress given by

$$\sigma_{ij} = -\frac{1}{V} P_{ij}.$$

The measure of the stress in such simulations gives thus a direct access to the value of the force moments.

Before applying the above method to our potential, one can notice that due to the symmetry of the octahedral interstitial site in the bcc lattice, the force moment tensor corresponding to a C atom takes the following expression in the reference frame corresponding to the cubic unit cell

$$(P_{ij}) = \begin{pmatrix} P_x & 0 & 0 \\ 0 & P_x & 0 \\ 0 & 0 & P_z \end{pmatrix},$$

for the [001] variant of the octahedral site. Other variants correspond to a permutation of the components $P_x$ and $P_z$. The stress tensor of a simulation box containing one C atom should therefore be diagonal with $\sigma_{xx} = \sigma_{yy}$.

The stress tensor measured in our simulations has the shape predicted by theory and its non-zero components vary linearly with the inverse of the simulation box volume (Fig. 3). The values deduced for the force moments from these simulations are $P_x = 3.40$ and $P_z = 8.03$ eV. One can compare these values with experimental data. Indeed, the tensor describing the strain induced by a C atom (Eq. 3) is diagonal too. This strain tensor therefore corresponds to a dilatation or contraction along the axes of the cubic unit cell. The fact that this strain varies linearly with the size of the simulation box for one carbon atom corresponds to Vegard law. This leads to the following linear relation between the variation of the lattice parameter and the atomic fraction $x_C$ of carbon atoms, all assumed to be in the [001] variant of the octahedral sites,

$$a(x_C) = a_0(1 + \delta_x x_C), \text{ along the [100] or [010] axes,}$$

$$c(x_C) = a_0(1 + \delta_z x_C), \text{ along the [001] axis,}$$

where $a_0 = 2.8553$ Å is the pure Fe lattice parameter as given by the atomic potential. The constants $\delta_x$ and $\delta_z$ corresponding to the previously deduced moments $P_x$ and $P_z$ are given by

$$\delta_x = \frac{2}{a_0^3} \frac{C_{11}P_x - C_{12}P_z}{(C_{11} - C_{12})(C_{11} + 2C_{12})} = -0.088,$$

$$\delta_z = \frac{2}{a_0^3} \frac{-2C_{12}P_x + (C_{11} + C_{12})P_z}{(C_{11} - C_{12})(C_{11} + 2C_{12})} = 0.56.$$

where we used the Fe experimental elastic constants $C_{11} = 243$, $C_{12} = 145$ and $C_{44} = 116$ GPa, the atomic potential used in this study being fitted on these constants [24,25].

Using the same potential, Becquart *et al.* [13] determined with molecular dynamics simulations the variation of Fe lattice parameter with C content at 300 K. They obtained $\delta_x = -0.1$ and $\delta_z = 0.6$. This gives us confidence in our method to determine the variations $\delta_x$ and $\delta_z$ of the lattice parameter and thus the force moments $P_{ij}$ from the stress. The difference between both results should arise from the temperature dependence of the coefficients $\delta_x$ and $\delta_z$.

All the authors that modelled the binding energy between C atoms and dislocation in iron within elasticity theory used different values for the coefficients $\delta_x$ and $\delta_z$. Douthwaite and Evans [19] used $\delta_x = -0.07$ and $\delta_z = 0.83$, these values being deduced from experimental measurements (dilatometry and anelasticity) in ferrite. On the other hand, Cochardt *et al.* [18] and Bacon [38] used values deduced from experimental measurements performed on martensite: $\delta_x = -0.052$ and $\delta_z = 0.76$ for Ref. [18], $\delta_x = -0.0977$ and $\delta_z = 0.862$ for Ref. [38]. Finally, Cheng et al. [39] obtained from a fit to different experimental data $\delta_x = -0.09$ and $\delta_z = 0.85$. Therefore, we see that the potential used in this study leads to a volume of formation $(2\delta_x + \delta_z)a_0^3/2$ and a tetragonal distortion $(\delta_z - \delta_x)$ smaller than the values used by all other studies based on elasticity theory. As the purpose of this article is mainly to compare atomic simulations with elasticity theory, we use the values $\delta_x = -0.088$ and $\delta_z = 0.56$ corresponding to the atomic potential.

## II.2 Point defect interaction with dislocation

A point-defect modelled by the force moment tensor $P_{ij}$ interacts with the strain field $\varepsilon_{ij}^d$ of the dislocation. The binding energy, as defined at the atomic scale by Eq. 1, is given by

$$E^{bind} = P_{ij}\varepsilon_{ij}^d. \qquad (Eq.\ 4)$$

Using Eq.3, this interaction energy can also be written

$$E^{bind} = V\varepsilon_{ij}\sigma_{ij}^d, \qquad (Eq.\ 5)$$

where $\sigma_{ij}^d = C_{ijkl}\varepsilon_{kl}^d$ is the stress created by the dislocation and $\varepsilon_{ij}$ is the homogeneous strain induced by the point defect on the volume $V$ as described in the previous subsection. This equation corresponds to the model first developed by Cochardt *et al.* [18] and used in all the other studies based on elasticity theory [19, 38]. Bacon [38] already pointed that Cochardt's model could be rationalized in term of force dipoles, leading to the equivalence between Eq. 4 and Eq. 5 for the binding energy.

It is sometimes considered that the point-defect is only a dilatation centre. This means that the tensors $P_{ij}$ or $\varepsilon_{ij}$ are diagonal with all diagonal components being equal. One thus obtains the elastic model first proposed by Cottrell and Bilby [1] and the binding energy reduces to the size interaction,

$$E^{\text{bind}} = -P^d \delta\Omega,$$ (eq.6)

where $P^d = -\sum_i \sigma_{ii}^d / 3$ is the pressure created by the dislocation and $\delta\Omega = V\sum_i \varepsilon_{ii}$ is the point defect relaxation volume. For a point defect like a vacancy which can truly be assumed to act as a dilatation centre (at least in crystals having the cubic symmetry), this approximation is correct as was shown in fcc metals [20]. This should be true too for substitutional impurities, but for interstitial impurities like C atoms in iron, this approximation is wrong as we will show below.

Studies of point-defect interaction with dislocations based on elasticity theory differ on the way they consider anisotropy too: one can use either isotropic, like Cochardt *et al.* [18] and Bacon [38], or anisotropic elasticity, like Douthwaite and Evans [19], to calculate the strain and stress created by the dislocation. It should be pointed out that when taking into account anisotropy in elasticity theory, the stress and strain created by the dislocation is still decaying as the inverse of the distance to the dislocation line. The point defect interaction energy with the dislocation will therefore observe the same dependence with the separation distance considering or not anisotropy. Only the angular dependence and the amplitude will differ.

For the anisotropic elastic calculations presented in this work, elastic constants corresponding to the Fe potential [24,25] are used, $C_{11}$ = 243, $C_{12}$ = 145 and $C_{44}$ = 116 GPa. For isotropic elastic calculations, we need to define equivalent isotropic elastic constants. To do so, we used the shear modulus $\mu$, the bulk modulus $K$ and the Poisson coefficient $\nu$, obtained by Voigt average [16] of the real anisotropic elastic constants,

$$\mu = \frac{1}{5}(C_{11} - C_{12} + 3C_{44}) = 89.2 \text{ GPa},$$

$$K = \frac{1}{3}(C_{11} + 2C_{12}) = 178 \text{ GPa},$$

$$\nu = \frac{C_{11} + 4C_{12} - 2C_{44}}{2(2C_{11} + 3C_{12} + C_{44})} = 0.285.$$

The use of this definition for the isotropic elastic constants has already been shown to lead to correct results when computing the vacancy – dislocation interaction energy in fcc metals [20].

## II.3 Comparison with atomic simulations

In figures 4-7, we plot the variation of the binding energy between a carbon atom and a dislocation, either screw or edge. Each sub-figure corresponds to a different variant of the C interstitial. When representing the binding energy on these figures, the distance $h$ between the C atom and the dislocation glide plane is kept fixed while the projection $x$ on the glide plane of the separation distance between both defects is varied. These figures illustrate thus the variation of the binding energy when the dislocation is gliding and the first derivative of the plotted function will give the force exerted by the C atom on the gliding dislocation.

In figure 4, we compare the results of the atomic simulations with elasticity theory for the screw dislocation. This figure allows to conclude on the validity of the different approximations that can be made within elasticity to calculate this interaction. If anisotropic elasticity is used to get the dislocation stress field and if all the components of the elastic dipole representing the C atom are considered (size and shape interactions as given by Eq. 4 or **5**), a perfect quantitative agreement is obtained with the atomic simulations. If one uses isotropic elasticity instead, no such agreement is obtained although the variation of the binding energy remains correct qualitatively. Assuming now that the C atom only acts as a dilatation centre and thus considering only the size interaction (Eq. 6), the binding energy obtained from anisotropic elasticity completely disagrees with the ones deduced from our atomic simulations. In particular, it is clear that most of the interaction between the C atom and the screw dislocation arises from the tetragonal distortion induced by the interstitial and only a small part can be attributed to its dilatation.

Very similar conclusions can be drawn for the edge dislocation (Fig. 5). Anisotropic elasticity perfectly reproduces the variation of the binding energy whereas isotropic elasticity reproduces the global trends but do not leads to the same quantitative agreement. In both cases, one has to consider both the dilatation and the tetragonal distortion due to the interstitial. Because of the higher pressure created by the edge dislocation than by the screw, the part of the binding energy associated with the dilatation is more important. Nevertheless, if one considers only this contribution and neglects the interstitial interaction with the shear created by the dislocation, no good description of the binding energy can be obtained.

This comparison (Fig. 4 and 5) between atomic simulations and elasticity theory has been made for a C atom lying far enough from the dislocation glide plane so as to be sure that elasticity could be applied. It showed that no approximation in the elastic calculations could be made to obtain quantitative predictions. We can now compare our atomic simulation results with the elastic calculations when the C atom gets closer to the dislocation glide plane (Fig. 6 and 7), so as to see what is the minimal separation distance for elasticity to apply. For a screw dislocation (Fig. 6), elasticity theory still manages to predict quantitatively the binding energy even when the C atom is really close to the dislocation centre. A discrepancy is observed only for positions which are at a distance smaller than 2 Å from the centre and which can be considered as belonging to the dislocation core. For all other positions, calculations based on elasticity theory will lead to the same binding energy as the atomic simulations.

The conclusions are quite different for a C atom interacting with an edge dislocation (Fig. 7). When the C atom gets closer to the dislocation glide plane the agreement between elasticity theory and the atomic calculation breaks up more rapidly than with the screw dislocation. It appears that elasticity theory perfectly reproduces the atomic simulations only when the C atom is further than ~20 Å, a region which cannot be reasonably assumed to correspond to the dislocation core.

**III. Discussion**

Two different hypotheses can be made so as to explain why the C atom has to be further from the edge dislocation core than from the screw one for elasticity to quantitatively predict the interaction energy.

Screw and edge dislocations create different stress fields. The screw stress field is mainly of shear type whereas the edge stress field has a strong hydrostatic and shear component. Indeed, elasticity theory predicts that both the pressure and the Von-Misès equivalent shear stress created by a Volterra edge dislocation reach a maximum of 40 GPa in the (110) plane which is the closest to the glide plane ($h = d_{110}/2 \approx 1$ Å). For a Volterra screw dislocation, the Von-Misès stress reaches the same value for the same separation distance but the maximum of the pressure is only 7 GPa. Therefore, the pressure component of the stress field is higher for the edge dislocation than for the screw dislocation. One thus expects the equation assuming linearity between the carbon interaction energy and the stress where it is embedded (Eq. 4) to be less precise. Indeed, it was shown in a recent study using the same potential [21] that the behaviour of the carbon diffusion barrier under a uniaxial stress is non-linear. This non-linearity is negligible for small stresses but have a strong influence when the stress increases like close to the dislocation core. One could consider this non-linearity to model within elasticity theory the carbon-dislocation interaction energy and obtain a better agreement with the atomic simulations. Actually, Eshelby theory for the elastic inclusion and inhomogeneity [40,41] allows to predict such a non-linearity leading to a correction which varies quadratically with the stress. Equivalently, one could consider that the amplitude of the elastic dipole used to model the point defect depends linearly on the local stress, thus introducing polarizability. In the case of a dilatation centre like a vacancy, one can obtain easy-to-use analytical expressions within isotropic elasticity. It has been shown that taking into account this second order correction slightly improves the agreement with atomic simulations for the vacancy binding energy to a dislocation [20]. For a point-defect with a tetragonal distortion, like C interstitial, an analytical expression of this second order correction is still tractable within isotropic elasticity [41]. Nevertheless, we have shown above that one needs to take full account of anisotropy in the elastic calculations to obtain a quantitative agreement with the atomic simulations at least for long separation distances. Such a second order correction would therefore have to be considered within anisotropic elasticity and would require the determination of the full 4$^{th}$ rank polarizability tensor of the point defect, which is far beyond the scope of this article.

Another possible explanation of the discrepancy at short distances between the elastic and atomic calculations of the binding energy for the edge dislocation is the displacement field created by the dislocation. If one compares the dislocation displacement field given by the atomic calculations using Mendelev potential with the Volterra displacement field predicted by anisotropic elasticity (Fig. 8), one sees that they disagree at short distances below ~20 Å. This is particularly true for the longitudinal component, *i.e.* the component parallel to the dislocation direction. At long distances, the atomic simulations lead to a displacement field corresponding to the Volterra one, but at short distances there is a supplementary displacement. It has been shown that a dilatation of the dislocation core can create such a supplementary displacement field that will superimpose the Volterra one [42]. This elastic field is much shorter range than the Volterra one, as the corresponding displacement is decaying as the inverse of the distance to the dislocation. It could explain why the elastic calculations lead to a value slightly different from the atomic calculations close to the dislocations. Indeed, our elasticity model assumes that a dislocation only creates a Volterra elastic field and does not take into account any other possible elastic field. For the screw dislocation, our atomic calculations lead to a displacement field that perfectly agrees with the Volterra elastic field and no such supplementary field is present with Mendelev potential. Therefore, it is not surprising that a better agreement between the atomic calculations and elasticity theory is observed for the screw than for the edge dislocation.

**Conclusions**

The binding energy in iron between a carbon atom and a dislocation has been studied at the atomic scale with an empirical interatomic potential and molecular statics simulations. The binding energy was found to be $E^{\text{bind}} = 0.66$ eV for an edge dislocation and $E^{\text{bind}} = 0.41$ eV for a screw dislocation, in reasonable agreement with available experimental data and results from other atomic simulations.

The atomic simulations have been then used to check the ability of elasticity theory to predict this binding energy. It has been shown that, to be quantitative, elasticity theory does not suffer any approximation. A perfect agreement with the atomic simulations is obtained when both the dilatation and the tetragonal distortion due to the C atom are considered and when anisotropy is included in the elastic calculations. Using isotropic elasticity theory instead, one can reproduce only qualitatively the interaction of the dislocation with the C atom.

For the screw dislocation, the binding energy predicted by elasticity theory is in very good agreement with the atomic simulations even when the carbon atom is close to the dislocation core. Some discrepancies exist as could be expected in the dislocation core. However, one can consider that the agreement is almost perfect for all the octahedral sites situated at a distance larger than 2 Å. For the edge dislocation, the picture is quite different. Far from the dislocation, the same quantitative agreement between elasticity and the atomic simulations is observed. But the elastic predictions differ from the atomic results when the distance between the C atom and the dislocation centre is smaller than 20 Å: both methods leads to the same trends for the variation of the binding energy but it was not possible to obtain an agreement as good as for the screw dislocation. Two different possibilities to explain this difference between screw and edge dislocations have been proposed. The polarizability of C atoms may have to be considered in the case of the edge dislocation because of the stresses which get higher than for the screw dislocation. A short range elastic field has also been evidenced close to the edge dislocation core and it may have to be considered in addition to the Volterra elastic field when computing the C atom interaction with the edge dislocation.

**Acknowledgments**

The authors are grateful to C. Domain, G. Monnet, L. Ventelon and F. Willaime for useful discussions on screw dislocations in iron, as well as to J.-L. Bocquet for his careful reading of the manuscript and fruitful discussions on elasticity theory.

*Table I: Maximum C-dislocation binding energy (eV) (Eq. 1). Experimental data have been obtained by anelastic measurements (Snoek damping peak and cold-work damping peak). Theoretical values based on elasticity theory used r = b as the cut-off distance.*

| Dislocation type | Authors | Method | $E^{bind}$ (eV) |
| --- | --- | --- | --- |
| Screw | Present work | MS | 0.41 |
| Screw | Chang [11] | MS | 0.59 |
| Screw | Cochardt et al. [18] | Isotropic elasticity | 0.75 |
| Screw | Douthwaite and Evans [19] | Anisotropic elasticity | 0.621 |
| Edge | Present work | MS | 0.66 |
| Edge | De Hosson [12] | MS | 0.7 |
| Edge | Tapasa et al. [14] | MS | 0.68 |
| Edge | Shu and Wang [15] | MS | 0.78 |
| Edge | Cochardt et al. [18] | Isotropic elasticity | 0.75 |
| Unknown | Kamber et al. [5] | Exp. | 0.5 |
| Unknown | Gavril'yuk et al. [43] | Exp. | 0.75 |
| Unknown | Henderson [44] | Exp. | 0.45 |

**Figures**

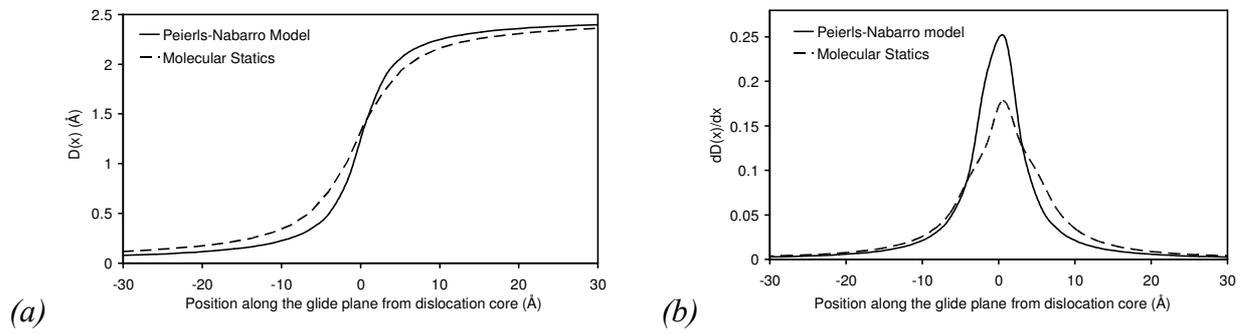

*Figure 1: Detection of the edge dislocation: a) the disregistry function as predicted by the Peierls-Nabarro model and given by molecular statics simulations, and b) the corresponding Burgers vector density distribution.*

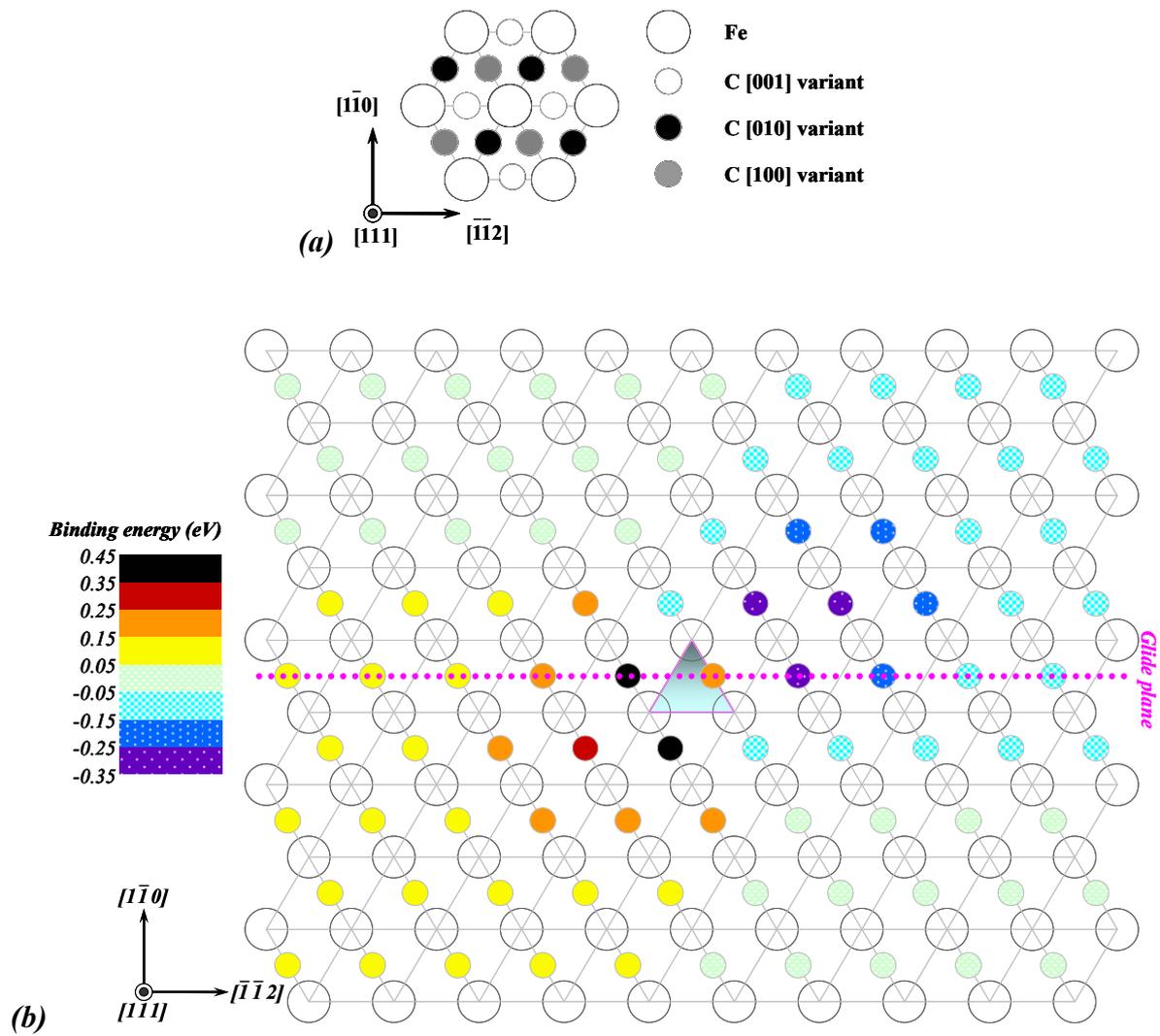

*Figure 2: (a) Projection on (111) plane of Fe atomic positions and C octahedral sites. (b) C-screw dislocation binding energies for different C [010] variant. The octahedral sites are coloured following a scheme depending on their binding energy. The position of the dislocation corresponds to the gravity centre of the coloured triangle. Binding energies for other C variants can be simply obtained by a rotation of ± 2π/3. [See the electronic edition of the Journal for a colour version of this figure.]*

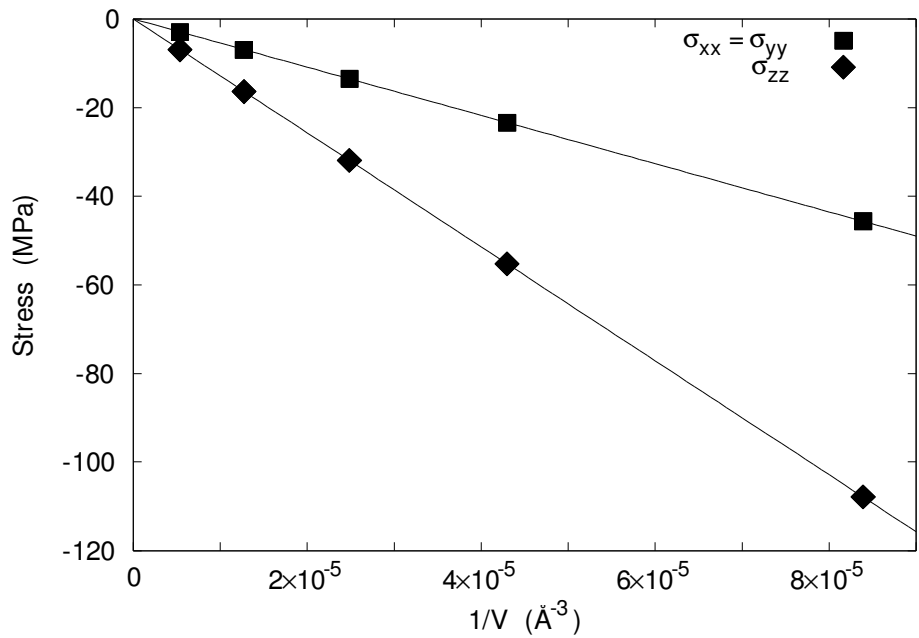

Figure 3: *Variation of the stress with the inverse of the simulation box volume for a simulation box containing one C atom in a [001] octahedral site. The stress is expressed in the reference frame corresponding to the cubic unit cell. Symbols correspond to atomic simulations and lines to their linear regression.*

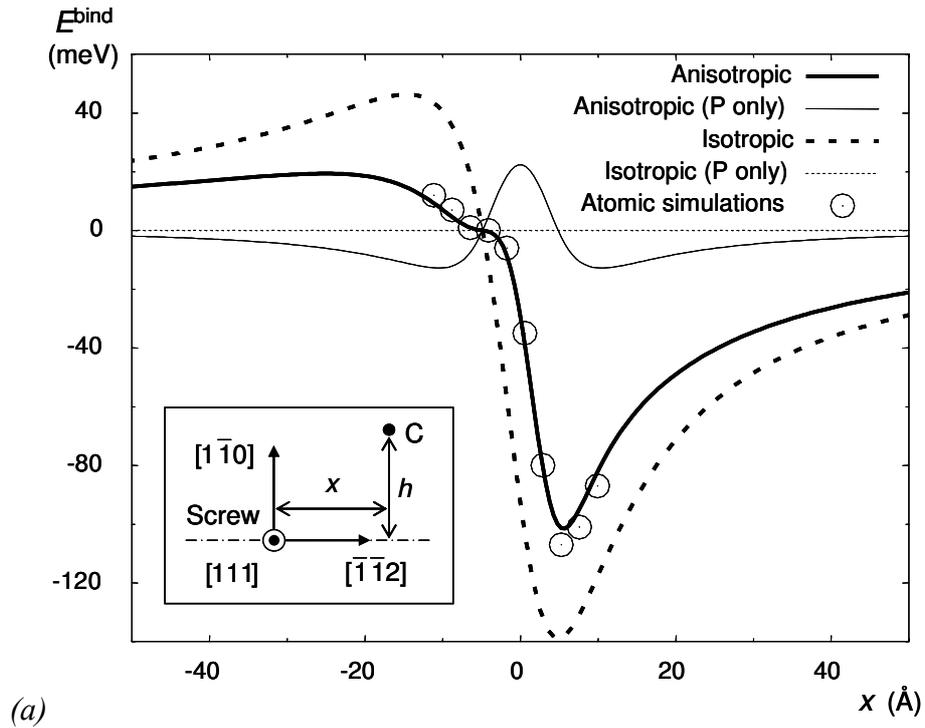

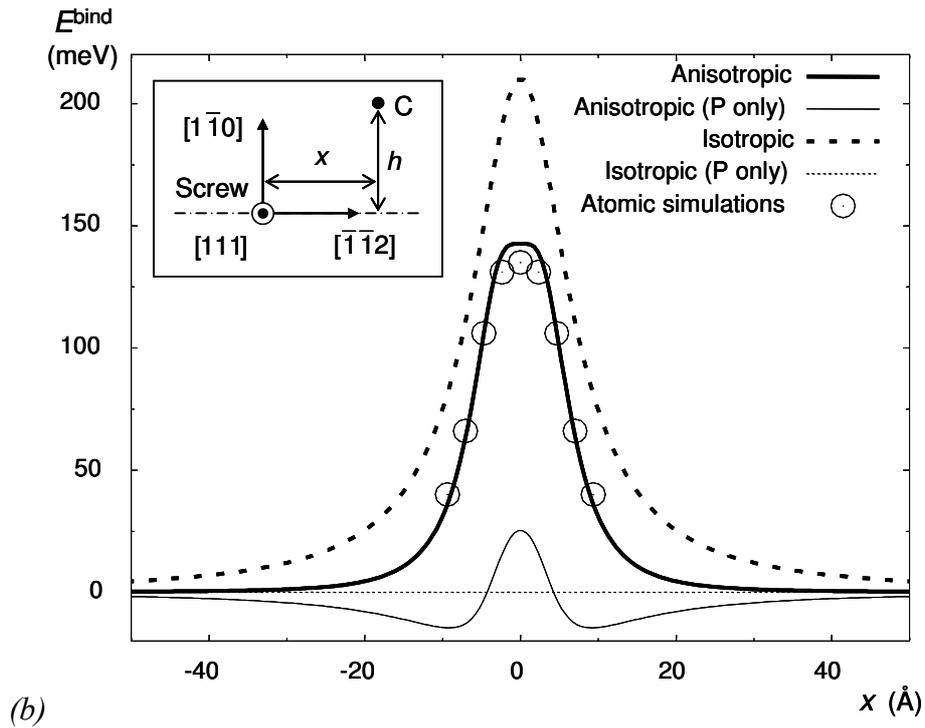

Figure 4: *Binding energy between a* $1/2[111](1\bar{1}0)$ *screw dislocation and a carbon atom for different positions x of the dislocation in its glide plane. The C atom lies (a) in a [100] octahedral site in the plane* $h = 4d_{110} \approx 8.1 Å$ *above the glide plane, (b) in a [001] octahedral site in the plane* $h = 3.5d_{110} \approx 7.1 Å$. *Symbols correspond to atomic simulations and lines to elasticity theory, considering all components of the stress created by the dislocation or only the pressure and using isotropic or anisotropic elasticity to calculate this stress.*

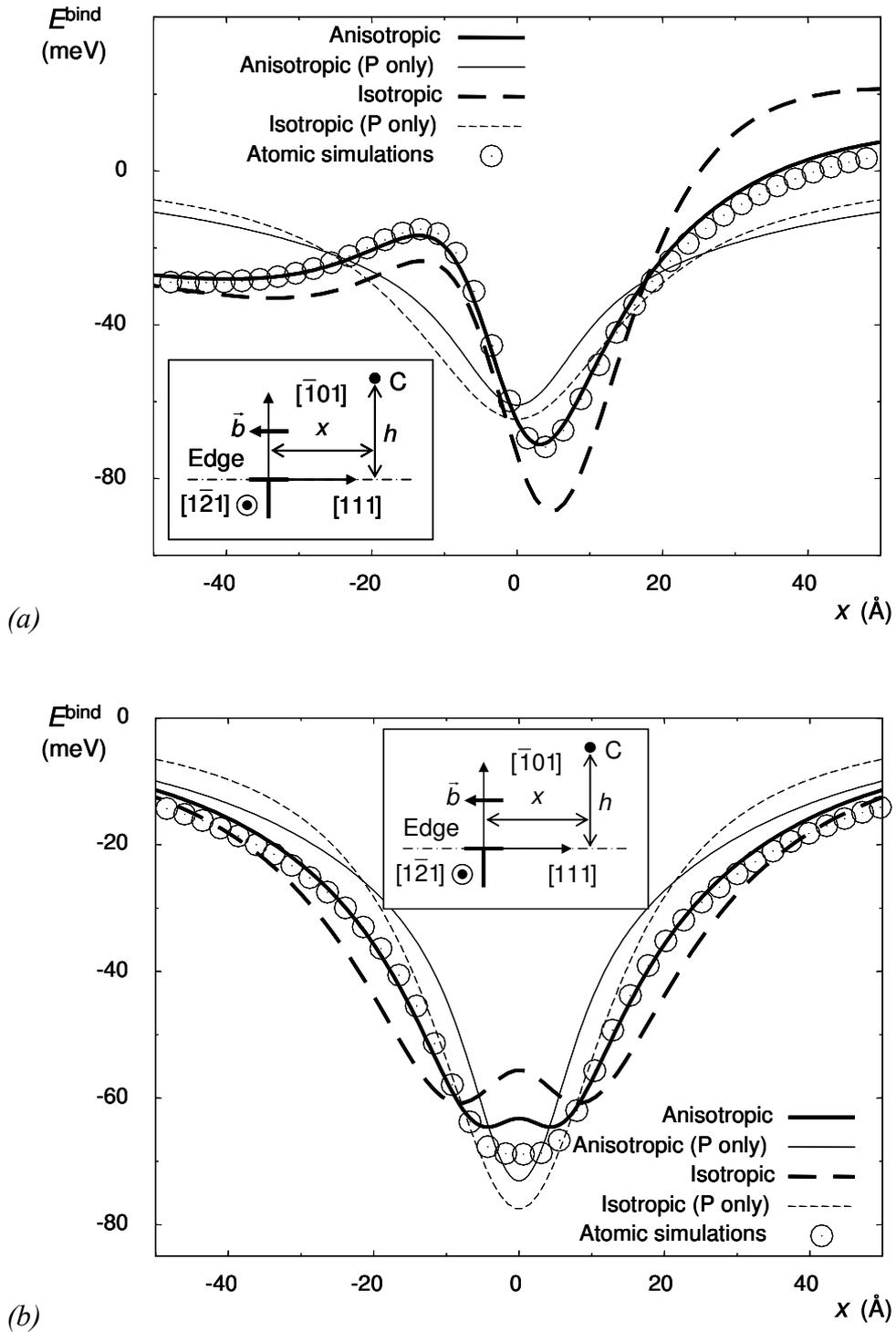

Figure 5: *Binding energy between a $1/2[111](\bar{1}01)$ edge dislocation and a carbon atom for different positions x of the dislocation in its glide plane. The C atom lies (a) in a [100] octahedral site in the plane $h = -9d_{110} \approx -18.2\,\text{Å}$ below the glide plane, (b) in a [010] octahedral site in the plane $h = -7.5d_{110} \approx -15.1\,\text{Å}$. Symbols correspond to atomic simulations and lines to elasticity theory, considering all components of the stress created by the dislocation or only the pressure and using isotropic or anisotropic elasticity to calculate this stress.*

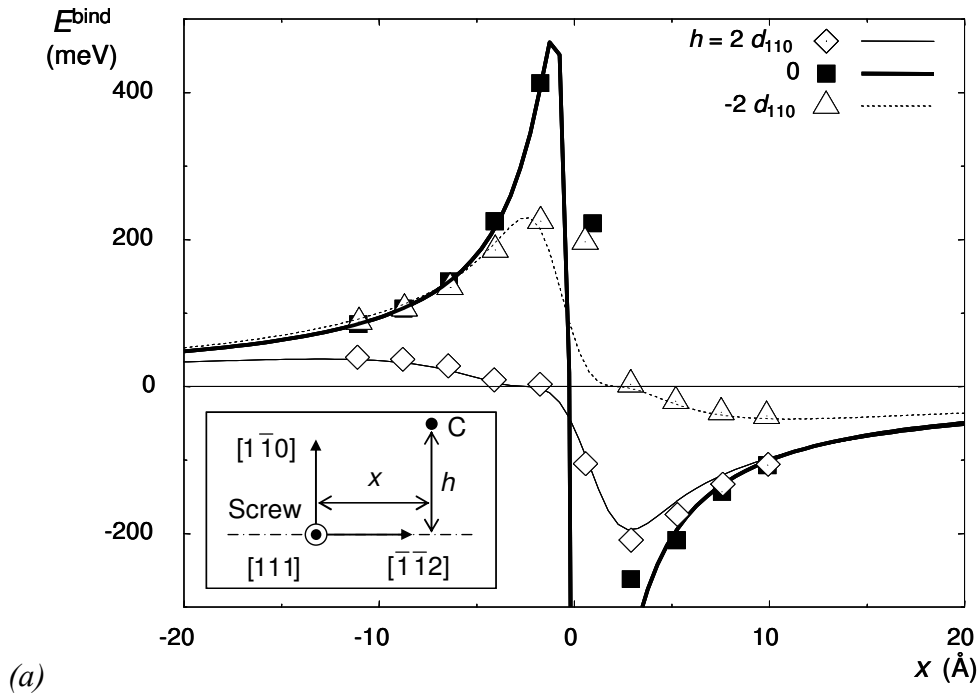

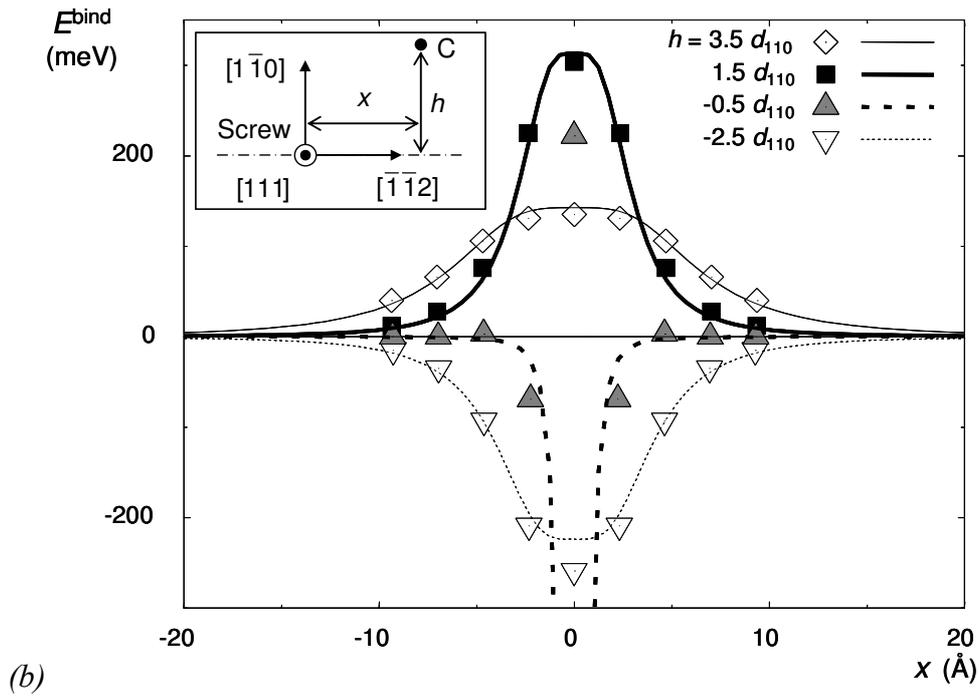

Figure 6: *Binding energy between a* $1/2[111](1\bar{1}0)$ *screw dislocation and a carbon atom for different positions x of the dislocation in its glide plane. The C atom lies (a) in a [100] octahedral site, (b) in a [001] octahedral site. Symbols correspond to atomic simulations and lines to anisotropic elasticity theory, considering all components of the stress created by the dislocation.*

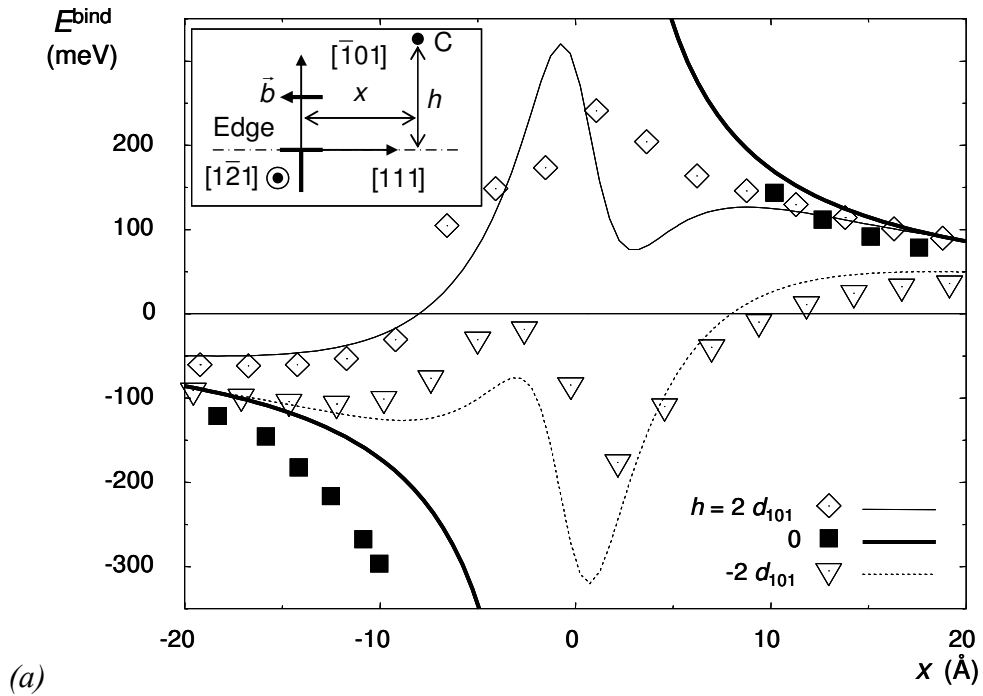

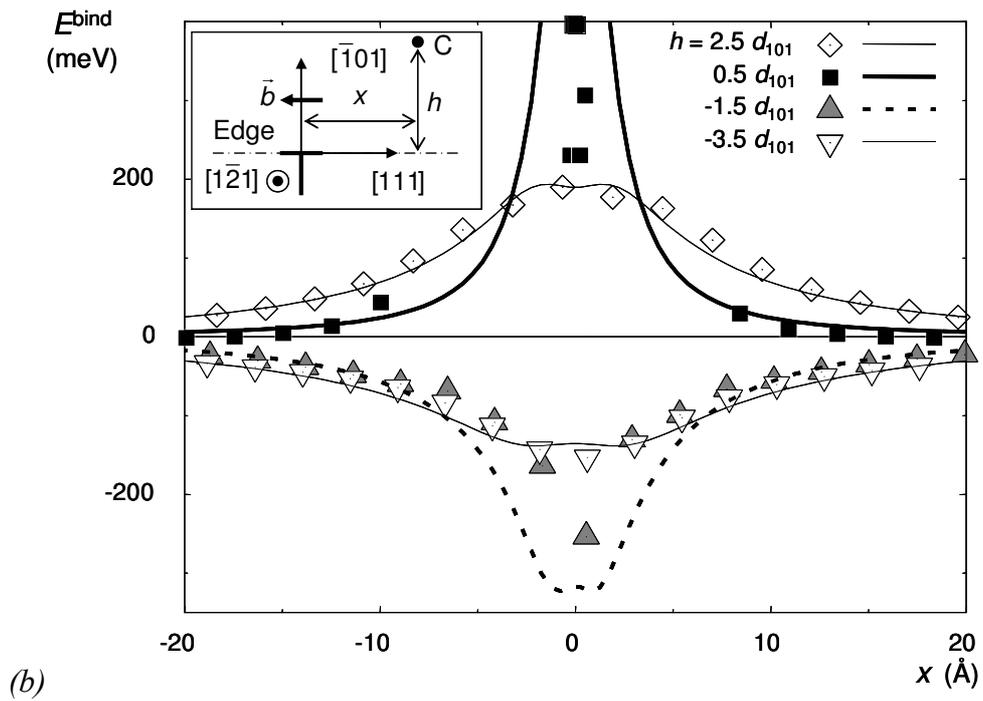

Figure 7: *Binding energy between a* $1/2[111](\bar{1}01)$ *edge dislocation and a carbon atom for different positions x of the dislocation in its glide plane. The C atom lies (a) in a [100] octahedral site, (b) in a [010] octahedral site. Symbols correspond to atomic simulations and lines to anisotropic elasticity theory, considering all components of the stress created by the dislocation.*

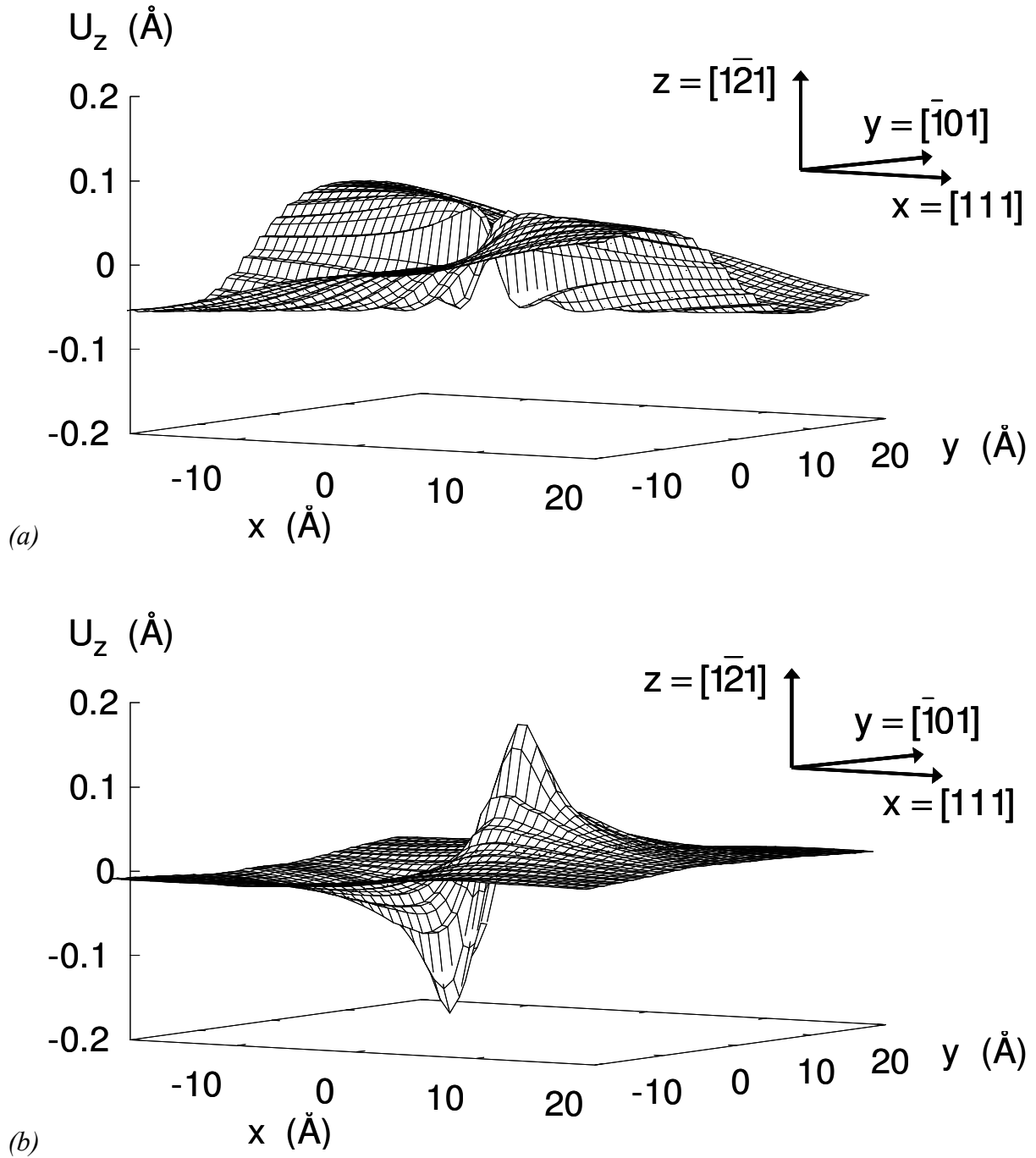

Figure 8: *Longitudinal component of the displacement field for an edge dislocation in Fe: (a) is the Volterra displacement field given by anisotropic elasticity. (b) is the difference between the displacement field observed in atomic simulations with Mendelev potential [24,25] and the Volterra displacement field given by anisotropic elasticity.*